# Distance Is Not Dead:
# Social Interaction and Geographical Distance in the Internet Era

Jacob Goldenberg[1] and Moshe Levy[2]

[1]The Hebrew University, Jerusalem 91905, Israel

[2] The Hebrew University, Jerusalem 91905, Israel

## ABSTRACT

The Internet revolution has made long-distance communication dramatically faster, easier, and cheaper than ever before. This, it has been argued, has decreased the importance of geographic proximity in social interactions, transforming our world into a "global village" with a "borderless society". We argue for the opposite: while technology has undoubtedly increased the overall level of communication, this increase has been most pronounced for *local* social ties. We show that the volume of electronic communications is inversely proportional to geographic distance, following a Power Law. We directly study the importance of physical proximity in social interactions by analyzing the spatial dissemination of new baby names. Counter-intuitively, and in line with the above argument, the importance of geographic proximity has dramatically *increased* with the internet revolution.

Acknowledgments

We are very grateful to Shirly Bitansky, Assaf Hefler, Yan Keren, Tal Levy, Abigail Rakover, Greg Ravikovich, Rom Schrift, Rachel Vovnoboy, and Jacob Shapiro, for their assistance and many helpful comments. We are grateful to Jonah Berger, Haim Levy, Barak Libai, Oded Netzer, Eran Rubin, and Olivier Toubia for their helpful comments and suggestions. This research has been financially supported by the Israel Science Foundation, and by the Davidson, K-mart and Zagagi Funds.

## Introduction

Information Technology (IT) allows us to do things that were undreamt of only a few decades ago. Today, as we sit in our favorite coffee shop in Manhattan, we can chat, view, and exchange files, or play chess with our friends anywhere across the globe. Common wisdom is that the IT revolution has reduced the importance of geographical proximity, creating a "borderless society," and a "global village" (Green, and Ruhleder, 1995;Farazmand, 1999). It also makes shortcutting ties between clusters leading to a dramatic increase of the shortcuts in the "small world" typical social network structure (see Watts, and Strogatz, 1998). We argue that the opposite is the case: in our contemporary IT-intensive world, geographical proximity has become an even stronger force than ever before.

Far before IT attracted attention, McLuhan and Fiore (1967) predicted that the development and integration of electronic technologies would lead to the emergence of a global village. In 1992, before the explosion of the IT revolution, when the Internet was still mainly used by universities and research institutes, Cleveland already argued that traditional notions "ownership" and "control" will have to change as local resources, markets and physical proximity will lose their importance. Many others have argued that these technologies will make it possible to think of the world as a single global village.

Information technology is a key element in achieving these purposes. In fact, these technologies may serve to define a whole new paradigm of organizations: "The borderless economy discussed here is really a whole information society arising out of the spread of new computer/communications technologies" (Dobell, and Steenkamp, 1993). But since the vision of the IT revolution was formed before the transformation actually occurred, could it be that its affect on the borderless society view is a myth, or is it a real fact?

The rationale of a borderless society appears to be sound. Research on social networks consistently shows that human societies tend to organize themselves in a "small world" structure (Watts, and Strogatz, 1998) a scale free structure (i.e., consisting of hubs, see (Barabasi, and Albert, 1999;Albert, et al., 2000) that lead to high clustering, and very short, efficient paths of communication. In 1967, Milgram's well-known experiment tested the hypothesis that members of any large social network (in his case, the population of the United States) are connected to each other through surprisingly short chains of intermediate acquaintances. His famous result is that the average length of the resulting acquaintance chains was roughly six, where the final member of the chain was the target itself. This result led to the phrase "six degrees of separation." Without going into the heated debate on whether Milgram's study supported his hypothesis (it is argued that only several dozen chains were ever completed), the finding that social structures facilitate fast communication through short paths is largely accepted. In their famous study, Dodds and Watts (2003) actually performed the first large scale, global verification of the small world hypothesis, using the modern Email equivalent of an earlier version of Milgram's experiment. The detailed analyses which included not only average acquaintance chain lengths, but also the distribution of lengths, showed that the world is indeed small, and as far as IT is concerned, access to information is effectively independent of geographical proximity. However, even Dodds and Watts' findings were inconclusive: Only 384 out of 24,163 chains were completed: the vast majority reached a "dead end," which may imply that even if the world has become smaller, borders may still exist.

On the other hand, there seems to be evidence that shows distance effects on network tie formation (Powell, et al., 2005). At the same time in many cases the fact that a network exists doesn't preclude, despite the easiness of communication, the importance of distance. For example, it has been shown that distance makes technological collaboration more difficult (Olson, and Olson, 2000;Kiesler, and Cummings, 2002).

In this paper, we study the effect of the IT revolution on social interactions. Our social interactions have a large part in defining who we are. Other than satisfying a basic human need, they influence what we think, what we buy, and how we invest (Hong, et al., 2005). Has physical proximity become a less important factor influencing our social interactions? There is no doubt that the IT revolution has intensified communications. *Long-distance* communications that were technically challenging and prohibitively expensive 20 years ago now entail simple, user friendly procedures that, in many cases, are almost costless. This seems to intuitively suggest that physical proximity now plays a more limited role in social interaction. However, while the effect of the new communication technologies on the long-distance communications is much more dramatic than its effect on local communications, we should keep in mind that most of our acquaintances are local. IT does not typically help us forge new acquaintances; it mostly helps us communicate with existing acquaintances. Most of us continue to establish new social relations in the traditional manner, through social activities and face to face meetings. Perhaps trust, friendship and bonding do not disseminate in the internet as easily as information. Thus, while the IT revolution increased the volume of all communications, it is possible that it has intensified our local communications to a greater extent than it intensified our global communications, simply because we maintain a greater number of local contacts. If this is correct, physical proximity may have become an even *more* important factor in social dynamics compared to the pre-IT era. This is the main issue we address in this paper.

We present two different studies to support the proposition that distance is plays an important role in social interaction even during the IT era. In the first study we describe the intensity of electronic communications as a function of physical proximity by examining the volume of email traffic as a function of geographical distance. We also investigate the density of social network contacts as a function of distance, using an extensive dataset of 100,000 Facebook network users. We show that both types of electronic social communication decrease inversely with the distance. Thus, we use electronic communications much more intensively locally, and this suggests that the IT revolution may have actually increased the importance of distance.

In the second study, we directly investigate the importance of physical proximity for social dynamics as a function of time by comparing the pre- and post- IT revolution eras. Using the Social Security dataset of baby names in the US, we trace the propagation of new baby names over the period extending from1970 to 2005. For the entire period we show that new names tend to diffuse geographically, suggesting that physical proximity is an important factor in the new name adoption dynamics. However, we show that the influence of physical proximity on baby name diffusion is much greater after the IT revolution occurred in the 1990s than before.

It does indeed seem, perhaps surprisingly, that physical proximity has become even more important for social dynamics in the IT era than before. The final part of the paper discusses the results and their implications.

# Study 1: The scale-free distribution of electronic communications

By definition, there is no data on the dependency of IT usage in social activity with geographical distance *before* the IT revolution took place. It is possible however to examine whether the geographical proximity indeed plays a prime role in social ties and communication. This is the purpose of study 1.

In this study we examine the relation between social interaction and physical distance. We analyze two sets of data: links of members on the Facebook social network, and email communications. In traditional forms of communications, physical distance typically affects the time, cost, and ease of communicating. Thus, data about traditional communication does not necessarily reflect the underlying distribution of social links: even if we have the same number of short-range links and long-range links, because of the higher cost and longer time required for long-distance communications we may communicate less with our long-range contacts. In contrast, for electronic communications physical distance is technologically irrelevant: it is just as easy to email someone who is 2000 miles away as it is easy to email our next-door neighbor. Thus, electronic communications provide an "uncontaminated" way to investigate the underlying distribution of social links. We study two types of social electronic communication: Facebook links and email.

Facebook is one of the largest and fastest-growing electronic social networks in existence today. In Facebook, users register, create personal profiles, and manage real-life and Internet friendships with other users.[1] Third parties are often invited to develop different applications for the Facebook platform. "My Personality" is such an application, which allows users to share their personal data, including their zip code. By examining pairs of friends (i.e. pairs of linked users) in the database, we constructed the distribution of link distances. The distance between any two linked users was calculated by converting each user's zip code to geographical longitude and latitude and calculating the distance. We collected data of 100,000 Facebook users, and found 1,297 linked pairs with reported zipcodes.

---

[1] See: http://www.facebook.com.

The distribution of Facebook links as a function of physical distance is shown in Figure 1.

Figure 1: The Distribution of Physical Distances of Facebook Contacts

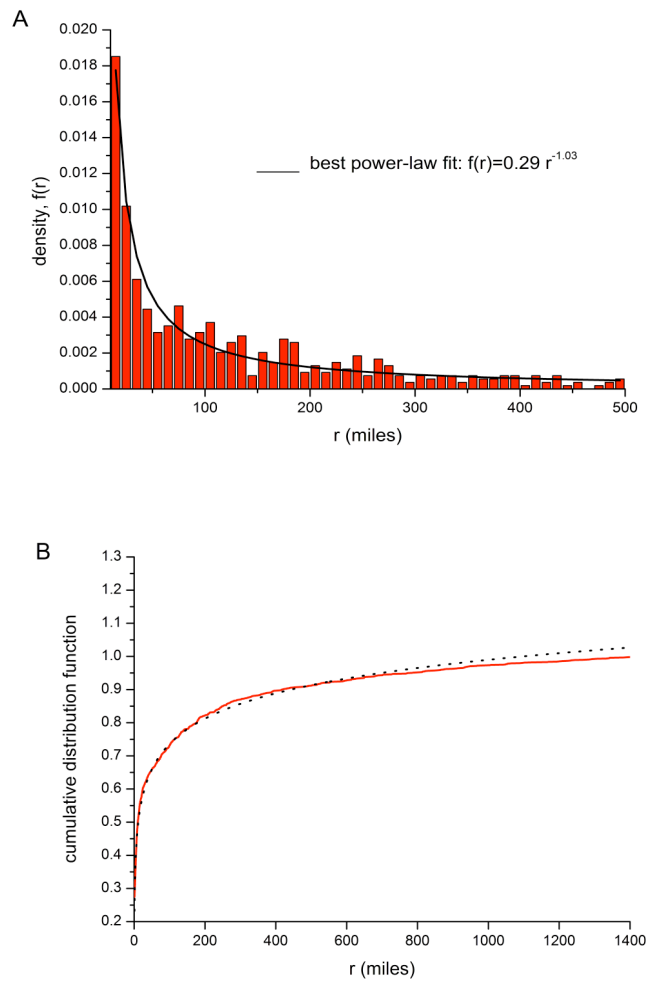

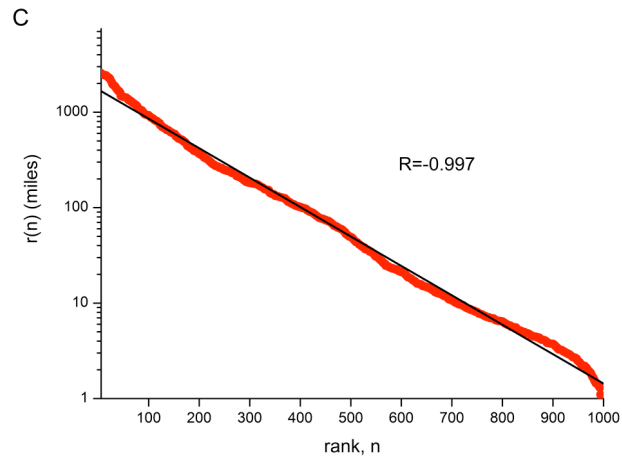

Panel A describes the empirical density function as a histogram. The empirical distribution is in very good agreement with a scale-free power-law distribution. The solid line shows the best power-law fit to the empirical data, with an exponent of -1.03 and a standard error of 0.03. Thus, the empirical distribution is in agreement with Zipf's (1949) Law, according to which density is proportional to *1/r*, where *r* is the distance (of course, for any finite-sized system, a *truncated* Zipf distribution must be considered). Panel B shows the empirical cumulative distribution (solid) and the best fit to the cumulative Zipf distribution (dashed), i.e. the best-fit logarithmic distribution (recall that the cumulative distribution for a truncated Zipf density function is the log function). While the density estimate provides a good picture for most distances, for large distances the number of observations is insufficient to obtain clear results using this method. Therefore, for large distances we also employ the rank-distance method, with findings presented in Panel C of Figure 1. In this method, we ranked the links from the one with the greatest distance (rank *n*=1) to the one with the lowest distance. The figure presents the distance *r(n)* as a function of rank, on a semi-logarithmic scale. The linear fit is again consistent with Zipf's Law: The linear relationship $log(r(n)) = A - Bn$ implies that the number of links exceeding distance *r* is $n(r) = \frac{A}{B} - \frac{log(r)}{B}$. If we denote the total number of links by *N*, the proportion of the distances greater than *r* is *n(r)/N*, and therefore the cumulative probability function *F(r)* is *1-n(r)/N*: $F(r) = 1 - \frac{A}{BN} + \frac{log(r)}{BN}$. The probability density function is the derivative of *F(r)* with respect to *r*, i.e. $f(r) = \frac{1}{BNr}$. Thus, a linear relationship between $log(r(n))$ and *n* implies Zipf's Law. The correlation we obtain between $log(r(n))$ and *n* is R = -0.997. To examine whether the obtained Zipf Law is a specific property of the Facebook network, below we employ a similar methodology to investigate the distribution of emails by distance.

Email is the most widespread form of electronic communications, with more than a billion users across the globe[2]. As it is extremely difficult to obtain detailed data on email communications, due to obvious privacy issues, little is known about the volume of email communications as a function of the geographical distance between correspondents. In this study we collected these data by asking subjects to report the locations of the recipients of their last 50 email messages, and their own city of residence (the complete questionnaire is provided in the supplementary material section). Overall, we collected data for 4,455 email messages.

The distribution of e-mail distances is reported in Figure 2.

---

2 An October 2007 report by the technology market research firm The Radicati Group (http://www.radicati.com/) estimates that there were 1.2 billion email users in 2007, and expects this number to rise to 1.6 billion by 2011.

Figure 2: The Distribution of Email Distances

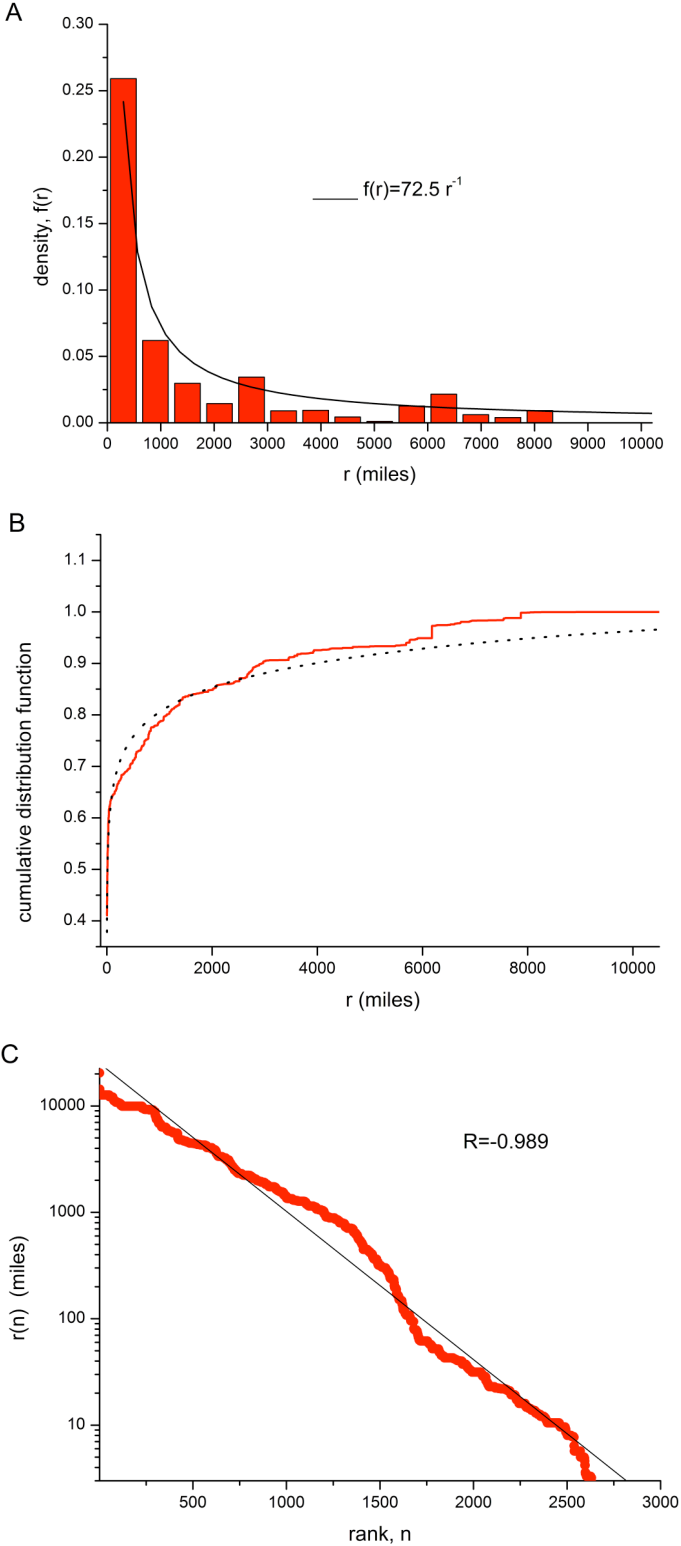

Panel A reports the density function. As the geographic location is given only at the city level of detail, the resolution of the email data is less detailed than the

Facebook data. Out of the 4,455 messages, 1818 (41%) where sent within the same city, yielding a distance measure of zero. Obviously, this low resolution limits our ability to characterize the density function, especially for short distances. As a result, Panel A of Figure 2 provides only a rough description of the density function, which appears consistent with Zipf's Law, but is not conclusive. Panel B provides a more detailed picture by presenting the cumulative distribution. The solid line describes the empirical cumulative distribution, while the dashed line shows the best logarithmic fit. The cumulative distribution is in good agreement with Zipf's Law. To complete the picture, Panel C describes the rank-distance relationship, with a correlation of R = -.989.[3]

It is striking that both email and Facebook communications depend on distance in a very similar way, and that this dependence is given by the simple Zipf Law, encountered in many other branches of science (Zipf, 1949;Mandelbrot, 1963;Egghe, 1991;Gabaix, 1999;Axtell, 2001;Li, and Yang, 2002;Yook, et al., 2002). A similar dependence has been found for internet routers (Lambiotte et. al. 2008)[4]. What does this regularity imply about the probability of two individuals, with a distance $r$ between them, being socially linked? Let us begin by making the simplifying assumption of identical "representative" individuals homogeneously spread over two-dimensional space. This implies that the number of individuals at distance $r$ from a given individual is proportional to $2\pi r$ (the circumference of a circle with radius $r$ around this individual). Denote the probability of a social link between two specific individuals with a distance r between them by *p(r)*, and the number of links of distance *r* that the individual will have by *f(r)*. *f(r)* is the number of "neighbors" at distance *r* multiplied by the probability of a link given this distance, i.e. *f(r)=2πr·p(r)*. As we empirically find *f(r)=c/r*, this implies that:

$$p(r) = \frac{c}{2\pi}\frac{1}{r^2}. \quad (1)$$

---

[3] Newman (2005) suggests an estimate of the power exponent based on log-likelihood maximization (see his eq.(5) on p. 327). While this estimate is generally superior to the slope estimates typically used in the literature, it is sensitive when the exponent value is close to 1 (see eq. B2 in his appendix). Employing Newman's method to our data sets yields exponent values slightly larger than 1: 1.12 and 1.20 for the facebook and email data, respectively.

[4] Liben-Nowell et. al. (2005) study the effect of distance in the LiveJournal network. They too find a power-law dependence, but they find that the link probability decreases as $r^{1.2}$.

Eq.(1), derived under the assumption of identical individuals, may be considered analogous to the gravitational law: the probability of a social link (the force) between two individuals (bodies of mass) is proportional to *1/r²*. Relaxing the assumption of identical individuals, by allowing individuals to have different susceptibilities (or unconditional probabilities) of being linked, $m_i$ and $m_j$, this analogy can be further extended to:

$$p(r) = \frac{Gm_i m_j}{r^2}, \quad (2)$$

where G is a constant given by $\frac{c}{2\pi\langle m\rangle^2}$.

Several scholars have predicted that electronic communications for which physical distance is completely irrelevant will fundamentally transform our social structure creating a "borderless society" (McLuhan, and Fiore, 1967;Ruhleder, and Green, 1993;Farazmand, 1999). Our empirical findings show that even though physical distance is technologically practically irrelevant for electronic communications, we use these communication paths primarily for short distances, because most of our social contacts are local.

This result implies that the IT revolution *may* have made distance even more important. Study 2 is designed to directly examine whether this is indeed the case.

## Study 2: Has Geographic Proximity Become More or Less Important? – Empirical Evidence from the Dissemination of New Baby Names

Our goal is to examine whether physical proximity has become more or less important for social interactions after the IT revolution in the 1990s. Such an investigation requires both spatial and temporal data about a social phenomenon that reflects social interactions, for a period spanning several decades, to cover the periods before and after the IT revolution. One such rare dataset is the detailed records maintained by the U.S. Social Security Service on new baby names.

When parents choose names for their new-born babies they are typically affected, at least to some degree, by their social interactions. This positive-feedback interaction can explain both the stellar climb up the popularity chart of several new "invented" baby names,[5] and also the fact that new baby names tend to spread spatially. Baby names offer a stringent test of our hypothesis because names are as equally communicable by phone or e-mail as by personal contact.

**The Data**

The Social Security Service publishes the list of the 1,000 most popular baby names in the US every year, and the number of babies given each of these names. In addition, the Social Security Service publishes the list of the 100 most popular baby names *in each state* in each year, including the number of babies given each name in that particular state. The data are available at: http://www.ssa.gov/OACT/babynames.

---

[5] A recent front-page story in the New York Times described the incredibly fast climb in the popularity of the baby girl name "Nevaeh" (Heaven spelled backwards). In 1999 only 8 newborn girls in the US were given this name. By 2005, Nevaeh became the 70$^{th}$ most popular name for newborn girls (ahead of Sara, Vanessa, and Amanda). In 2006, Nevaeh continued its stellar climb and ranked 43 on the popularity chart (becoming more popular than Allison and Maria). See "And if It's a Boy, Will it be Lleh?" by Jennifer Lee, *New York Times*, May 18, 2006. In 2008, Nevaeh ranked 34.

To analyze this dataset, we ran a crawler software application that allowed us to track the development of each name through time and across states. This software, which appears in the supplementary materials section, is available for public use.

**Results**

Our results are comprised of two parts. In the first part we show graphically that physical proximity plays an important role in the dissemination of new baby names. In the second part, we develop a formal measure of the importance of physical proximity to the dynamics, and track the behavior of this measure over the last 35 years.

**A. Physical Proximity is Important**

The importance of physical proximity in the dissemination of new baby names can be illustrated graphically by the dynamics of the name "Ashley" described in Figure 3.

Figure 3: The Dynamics of the Name "Ashley".

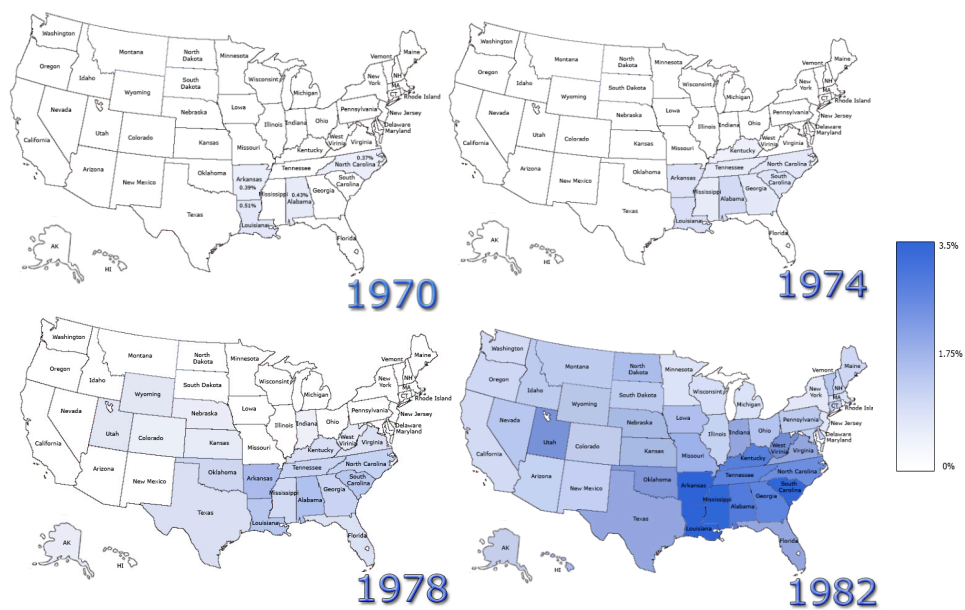

The first panel in the figure is a snapshot of the year 1970. In this year, the name "Ashley" appeared in the Top 100 name list in only four states: Alabama, Arkansas, Louisiana, and North Carolina. The second panel of Figure 1 shows a snapshot of "Ashley" for 1974 (for simplicity of the exposition, the percentage of Ashleys in each state out of the total number of new babies in the state is expressed by shade coding). The fact that "Ashley" spread to neighboring states between 1970 and 1974 suggests that physical proximity played an important role in the process of name adoption. The next two panels, representing 1978 and 1982, further support this notion.

Has physical proximity continued to play an important role after the IT communications revolution of the 1990s? While we address this question methodically in the next section, one example for this case is provided in Figure 4.

Figure 4: The Dynamics of the Name "Kaden"

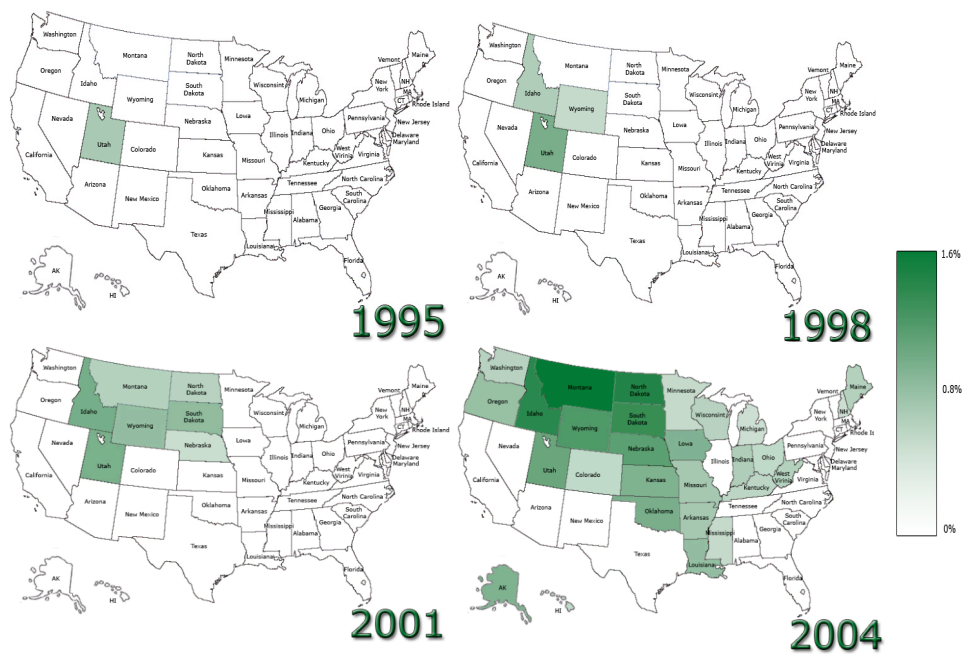

This figure describes the dynamics of the (boy) name "Kaden" between 1995 and 2003. The dynamics again suggest that physical proximity plays an important role. There is a clear similarity to the dissemination of an epidemic, with strong spatial correlations.

These are just two examples. The program in the supplementary materials section is available to examine the spatial development of any other name. The analysis in the next section employs the data of all names.

**B. How Has the Importance of Physical Proximity Changed Over Time?**

In order to address this question, we develop an index to measure the size of the physical proximity effect, and we then track the value of this index over time. Several different measures have been employed in previous research to capture the importance of proximity or effects in the diffusion of social phenomena (Mandelbrot, 1963). The measure we employ is in the same spirit, and is adopted to the context of baby-names. If physical proximity plays no role in the dynamics, all else equal, we would expect the proportion of new babies with name i to be distributed uniformly across the country. In contrast, if physical proximity does play a role, we expect the proportion of new babies with name i to be higher in the vicinity of the states where this name has already appeared. We separate states into two groups: Group $A^i(t)$: states in which the name i appeared on the Top 100 list before time t, plus the immediately bordering states; Group $B^i(t)$ is the group of all the remaining states. We employ the following notation:

$N_A^i(t)$: The number of babies with name i born in year t in Group A states.

$N_B^i(t)$: The number of babies with name i born in year t in Group B states.

$N_A^{total}(t)$: The total number of babies (all names) born in year t in the group A states.

$N_B^{total}(t)$: The total number of babies born in year t in the group B states.

If physical proximity plays no role in the diffusion of name i in the US, we expect the number of babies with name i born in Group A states to be

$$\left( \frac{N_A^{total}(t)}{N_A^{total}(t) + N_B^{total}(t)} \right) N^i(t), \tag{3}$$

The term in equation (3) reflects the expected proportion of babies named i in Group A states, in the absence of proximity effects, i.e., the proportion of i babies in Group A and Group B is equal. In contrast, if physical proximity *is* important, we expect $N_A^i(t)$, the number of babies named i born in Group A states, to be greater than the expression in (3). We define the Proximity-Effect Index (PEI) of name i at time t as follows:

$$PEI^i(t) \equiv \left[ \frac{N_A^i(t)}{\left( \frac{N_A^{total}(t)}{N_A^{total}(t) + N_B^{total}(t)} \right) N^i(t)} \right] - 1. \tag{4}$$

PEI measures the percentage by which the actual number of babies named i in Group A states exceeds the expected number of babies named i assuming no proximity effects. If, for example, $PEI^i(t) = 0.2$, this means that the number of babies named i born in Group A states exceeded the expected number of babies named i in the absence of proximity effects by 20%.

As we are interested in examining the overall trend in the proximity effect, we calculate the PEI for all the names. The development of the median PEI over time is depicted in Figure 5.[6] The fact that the median PEI is positive implies that physical proximity has played a role in the baby name dynamics over the entire 35-year period (as figures 3 and 4 suggest). In fact, over 99% of the PEI measures for individual names are positive. The typical value of approximately 0.2 for the median PEI in the period up to 1995 is rather large: this value implies that the proportion of babies with a particular name in states where this name appeared in the past and in their neighboring states is 20% higher than would be expected in the absence of proximity effects.

Figure 5: The Importance of Physical Proximity over Time- Baby Names.

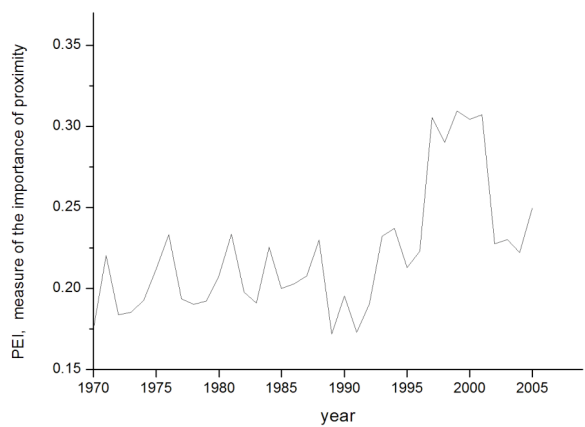

The most striking result revealed in Figure 5 is the fact that the median PEI increased dramatically over the decade of the IT revolution. While the importance of physical proximity was rather stable between 1970 and 1990, the communications revolution in the 1990s was associated with an increase in the importance of physical proximity. Between 1995 and 2002, the median PEI jumped to about 0.3, a 50% *increase* relative to the pre-revolution period. This surprising increase is in accord with the counter-intuitive argument made in this paper: While the communications revolution has no doubt made it easier to transmit information over great distances, its effect on local contacts may have been even more dramatic, enhancing proximity effects more than before. Although it declined somewhat after 2002, the median PEI level remained higher than typical pre-1990 levels.

---

[6] We use the median rather than the average PEI, because there are some large positive outliers in the PEI, driven by small states. For example, assume that a new name first appears in a small state with a small number of neighbors, or states outside the contiguous United States such as Hawaii, or Alaska. In this case, Group A comprises only that particular state, and $N^i(t) = N_A^i(t)$. Because the state is small, $\left(\dfrac{N_A^{total}(t)}{N_A^{total}(t) + N_B^{total}(t)}\right)$ is very small, and therefore $PEI^i$ can be extremely large. While the average PEI can be dramatically affected by such special cases, the median PEI is not similarly affected, and therefore is a much more robust measure. Similar results are obtained when different percentiles of the PEI are employed.

The exact point in time where the IT revolution took place is hard to define, and may be somewhat subjective. However, it is generally accepted that this revolution took place somewhere in the 1990's. If we take the middle of this decade as a reference point, we find that before 1995 the average value of the PEI shown is 0.203. In the post-1995 period the average value is 0.267. This difference is highly significant, with a t-value of 6.76. Thus, the baby-name data suggest that the IT revolution was accompanied by a significant *increase* in the importance of geographical proximity for social interaction.

## Discussion

The Information Technology revolution has dramatically changed our ability to communicate with each another. The new electronic communications options have rendered geographic proximity completely irrelevant by allowing us to transmit information across the globe instantaneously, with great ease. In the past, people believed that this technological revolution would transform society and social relations, leading to a borderless society in a "global village." These beliefs, however, were never empirically put to a test.

While the IT revolution has changed many important aspects of our lives, it seems that it has not fundamentally changed the structure of society and the organization of social relations. Empirical evidence suggests that contrary to expectations, geographical proximity has become *more* important for social interactions and dynamics than ever. We argued that one possible reason is that the major part of our electronic communications are performed with local counterparts. We demonstrated that while the IT revolution has clearly increased the overall volume of communications, it has increased local communications to a greater degree than long-distance communications: the volume of electronic communications as a function of the geographical distance follows Zip's inverse proportionality law for both Email and Facebook interactions.

The IT revolution may have transformed the world into a global village, but it has not changed us into a borderless society in which geographical proximity plays a more limited role in social relations. On the contrary, the importance of geographical proximity appears to have increased.